\documentclass[%
 reprint,
superscriptaddress,
 amsmath,amssymb,
 aps,
prstab,
]{revtex4-2}

\usepackage{graphicx}
\usepackage{dcolumn}
\usepackage{bm}

\usepackage{xcolor} 

\begin{document}


\title{Impact of coherent wiggler radiation impedance in Tau-Charm factories}

\author{Tianlong He}
\email{htlong@ustc.edu.cn}
\affiliation{%
 National Synchrotron Radiation Laboratory, University of Science and Technology of China, No. 42, South Cooperative Road, Hefei, 230029, China 
}%

\author{Ye Zou}
\email{zouye@ustc.edu.cn}
\affiliation{%
 School of Nuclear Science and Technology, University of Science and Technology of China, No. 443 Huangshan Road, Hefei, 230027, China 
}%

\author{Demin Zhou}
\email{dmzhou@post.kek.jp}
\affiliation{%
 KEK, 1-1 Oho, Tsukuba 305-0801, Ibaraki, Japan 
}%
\affiliation{
 The Graduate University for Advanced Studies, SOKENDAI
}%

\author{Hao Zhou}
\affiliation{%
 School of Nuclear Science and Technology, University of Science and Technology of China, No. 443 Huangshan Road, Hefei, 230027, China 
}%

\author{Hangzhou Li}
\affiliation{%
 National Synchrotron Radiation Laboratory, University of Science and Technology of China, No. 42, South Cooperative Road, Hefei, 230029, China 
}%

\author{Linhao Zhang}
\affiliation{%
 School of Nuclear Science and Technology, University of Science and Technology of China, No. 443 Huangshan Road, Hefei, 230027, China 
}%

\author{Tao Liu}
\affiliation{%
 School of Nuclear Science and Technology, University of Science and Technology of China, No. 443 Huangshan Road, Hefei, 230027, China 
}%

\author{Weiwei Li}
\affiliation{%
 National Synchrotron Radiation Laboratory, University of Science and Technology of China, No. 42, South Cooperative Road, Hefei, 230029, China 
}%

\author{Jingyu Tang}
\affiliation{%
 National Synchrotron Radiation Laboratory, University of Science and Technology of China, No. 42, South Cooperative Road, Hefei, 230029, China 
}%
\affiliation{%
 School of Nuclear Science and Technology, University of Science and Technology of China, No. 443 Huangshan Road, Hefei, 230027, China 
}%


\date{\today}

\begin{abstract}
Coherent synchrotron radiation (CSR) has long been recognized as a significant source of longitudinal impedance driving microwave instability in electron storage rings. In the pursuit of higher luminosity, next-generation circular $e^+e^-$ colliders operating in the few-GeV energy range, such as B-factories and Tau-Charm factories, are being designed with low-emittance beams and high beam currents. Damping wigglers are commonly introduced to reduce damping times and control beam emittance. In this study, we systematically investigate the impact of coherent wiggler radiation (CWR), a specific form of CSR generated within wigglers, on beam stability in Tau-Charm factories. We revisit the threshold conditions for CWR-induced microwave instability and evaluate its effects under realistic lattice configurations of collider rings. Furthermore, we examine theoretical models of longitudinal CWR impedance and identify improved formulations that better capture its influence. As an illustrative example, the developed CWR impedance models are applied to simulate beam stability in the Super Tau-Charm Facility currently under design in China.
\end{abstract}


\maketitle

\section{Introduction}\label{sec:introduction}

In particle accelerators, a charged beam emits synchrotron radiation (SR) as it traverses bending magnets. The total radiation spectrum can be expressed as  
\begin{equation}
P(k) = \bigl(N + N(N-1)|F(k)|^2\bigr)\, I_1(k),
\end{equation}
where $k$ is the radiation wavenumber, $N$ is the number of particles in the bunch, $F(k)$ is the bunch form factor, and $I_1(k)$ is the single-particle spectrum. The second term, proportional to $N^2 |F(k)|^2$, corresponds to coherent synchrotron radiation (CSR). This contribution becomes significant when the bunch length $\sigma_z$ is comparable to or smaller than the radiation wavelength $2\pi/k$, or when sub-structures with characteristic scale $\hat{\sigma}_z$ form within the bunch such that $F(k)$ remains sizable at $k \sim 1/\hat{\sigma}_z$. Hence, the coherence originates entirely from the bunch profile through its form factor.

In storage rings, the interaction of SR with the beam is usually described by the longitudinal impedance $Z(k)$, whose real part is directly related to the single-particle spectrum as $\mathrm{Re}[Z(k)] \propto I_1(k)$. By long-standing convention, this impedance is referred to as the ``CSR impedance''. Although this terminology can be somewhat misleading, since the coherence does not arise from the impedance itself, it is standard in the literature. In this paper, we follow this convention and use ``CSR impedance'' to denote the impedance of synchrotron radiation in bending magnets.

The influence of CSR fields on beam stability in storage rings has been the subject of extensive research in recent years (see, for example,~\cite{dastan2024coherent} and references therein). Of particular interest is the case of synchrotron radiation from wigglers, commonly termed coherent wiggler radiation (CWR). This effect was first identified in the 2000s as a potential driver of microwave instability during the design of damping rings for colliders~\cite{wu2003calculation, wu2003impact}. Following the terminology used for CSR, we refer to the associated impedance as the CWR impedance, while emphasizing that coherence originates from the bunch form factor rather than the impedance itself. More recently, the impact of CWR impedance on microwave instability has been investigated in detail through both theoretical analyses and numerical simulations~\cite{blednykh2023microwave}.

The crab-waist collision scheme has become the standard choice for new-generation circular $e^+e^-$ colliders, including B-factories~\cite{superb2007superb, SuperKEKBTDR}, Tau-Charm factories~\cite{biagini2013tau, anashin2018super, luo2019progress}, and Higgs factories~\cite{cepc2018cepc, fcc2019fcc}, since its invention in 2006~\cite{Raimondi2006} and its successful demonstration at DA$\Phi$NE~\cite{zobov2010test}. For collider rings operating in the few-GeV range (below 10 GeV), the radiation damping provided by regular bending magnets is insufficient to achieve the required damping time. As a result, damping wigglers are introduced to shorten the damping time and control the beam emittance (see, for example,~\cite{SuperKEKBTDR, biagini2013tau, anashin2018super}).


The use of damping wigglers, while essential, raises concerns about the impact of CWR-driven instabilities on beam dynamics in collider rings. In~\cite{wu2003calculation}, the CWR impedance in free space was evaluated, and later in~\cite{wu2003impact} the combined effects of CSR from bending magnets and wigglers were analyzed to optimize the performance of damping rings. A scaling law for the instability threshold as a function of the wiggler-to-bend damping ratio [Eq.~(24) in~\cite{wu2003impact}] was proposed to assess the role of CWR impedance. However, this scaling law relies on the plausible assumption that the momentum compaction factor $\alpha_p$ varies with the damping ratio, an assumption that does not hold in collider rings. Consequently, the threshold current scaling derived in~\cite{wu2003impact} is not directly applicable to the collider rings considered here.

Further progress was made in~\cite{stupakov2016analytical}, where an analytical framework incorporating parallel-plates shielding was developed for CWR impedance calculations. While this represents an important advance, the resulting formulas do not converge well in numerical evaluations, limiting their practical applicability. These considerations motivate us to establish a theoretical framework tailored to the specific requirements of modern collider rings, especially those employing the crab-waist collision scheme. Our approach also builds upon and extends recent advances in CSR and CWR instability studies~\cite{blednykh2023microwave, dastan2024coherent}. Moreover, although our analysis is carried out in the context of a Tau-Charm facility, the framework is general and can be applied to other machines, provided the applicability conditions are carefully examined.

The paper is organized as follows. In Sec.~\ref{sec:theories}, we briefly review the theories of longitudinal single-bunch instability in electron storage rings, with the CSR-induced instability as a specific case. Then we derive alternative analytic models for CWR impedance in Sec.~\ref{sec:cwr-impedance}. Using the theories, we evaluate the impact of CWR impedance on ring design for Tau-Charm factories in Sec.~\ref{sec:cwi}. In Sec.~\ref{sec:simulations}, detailed simulations are presented to analyze the microwave instability induced by the CWR and CSR impedances for the proposed Super Tau-Charm Facility (STCF) in China. Finally, we summarize our findings in Sec.~\ref{sec:summary}.

\section{Theories of longitudinal single-bunch instability}\label{sec:theories}

The longitudinal wakefields distort the potential well experienced by the beam in electron storage rings. When the bunch current exceeds a certain threshold, the beam becomes unstable. A widely used formula for estimating the threshold bunch current is provided by the Keil-Schnell-Boussard (KSB) criterion~\cite{boussard1975observation, wolski2023beam}:
\begin{equation}
    I_{th} (\text{KSB}) = \frac{(2\pi)^{3/2}(E/e)|\eta|\sigma_p^2\sigma_z}{C \left| \frac{Z}{n} \right|},
    \label{eq:boussard-criterion}
\end{equation}
where $E$ is the beam energy, $e$ is the electron charge, $\eta=\alpha_p-1/\gamma^2$ is the slip factor with $\alpha_p$ being the momentum compaction and $\gamma$ being the Lorentz factor, $\sigma_p$ is the energy spread, $\sigma_z$ is the Gaussian bunch length, $C$ is the ring circumference, and $\left| \frac{Z}{n} \right|$ is the longitudinal impedance per harmonic, a real quantity with units of impedance. For a given design bunch current, Eq.~\eqref{eq:boussard-criterion} also provides a guideline for managing the impedance level in a storage ring. Since we consider electron rings in the GeV energy range, we set $\eta = \alpha_p$ and tentatively assume $\alpha_p > 0$ as follows.


The quantity $\left| \frac{Z}{n} \right|$ is associated with the machine impedance, which can be determined in different ways. For longitudinal CSR impedances in the high-frequency range $ k \gg 1/\sigma_z $, the dispersion relation analysis for the coasting beam presented in~\cite{stupakov2002beam} was employed to determine the threshold current for microwave instability. Later, this method was extended to incorporate various analytical impedance models~\cite{cai2011theory, blednykh2023microwave, dastan2024coherent}. Furthermore, it was shown that the dispersion relation (e.g., Eq.~(9) in~\cite{stupakov2002beam}) can also be solved using numerically computed broadband impedances~\cite{zhou2011thesis, dastan2024coherent}. In particular, an explicit form of the frequency-dependent threshold current is found in~\cite{blednykh2023microwave}:
\begin{equation}
    I_{th}(k) = \frac{2\pi (E/e)\alpha_p\sigma_p^2\sigma_z k\text{Re}\left[Z_\parallel(k)\right]}{G_i \left| Z_\parallel(k) \right|^2},
    \label{eq:Ith-general}
\end{equation}
where $Z_\parallel(k)$ represents the longitudinal broadband impedance as a function of the wavenumber $k$ and $\text{Re}[]$ denotes the real part of a complex function. The factor $G_i$ is determined from~\cite{blednykh2023microwave}
\begin{equation}
    \frac{G_i(x)}{G_r(x)} =\frac{\text{Re}\left[Z_\parallel(k)\right]}{\text{Im}\left[Z_\parallel(k)\right]},
    \label{eq:GiGr_vs_ZrZi}
\end{equation}
at the given $k$, with
\begin{equation}
    G_r(x) =\sqrt{2\pi}-\pi x e^{-x^2/2} \text{erfi}[x/\sqrt{2}],
\end{equation}
\begin{equation}
    G_i(x) =\text{sgn}[\eta]\pi x e^{-x^2/2}.
\end{equation}
Here $\text{Im}[]$ denotes the imaginary part, $\text{erfi}[z]=-i\frac{2}{\sqrt{\pi}}\int_0^{iz} e^{-t^2}dt$ is the imaginary error function, and $\text{sgn}[]$ is the sign function. Both $G_r(x)$ and $G_i(x)$ are real-valued functions of the real variable $x$. Radiation damping is neglected in the derivation of Eq.~\eqref{eq:Ith-general}.


From the application of Eq.~\eqref{eq:Ith-general} to broadband high-frequency impedances, as demonstrated in~\cite{dastan2024coherent}, it has been empirically observed that the minimum threshold current occurs around a critical wavenumber $k_c$, which is close to the zero crossing of $\text{Im}[Z_\parallel(k)]$, i.e., $\text{Im}[Z_\parallel(k_c)] \approx 0$. Under this condition, Eq.~\eqref{eq:Ith-general} reduces to
\begin{equation}
    I_{th}(k_c) = \text{Min} \left[ I_{th}(k) \right] 
    \approx \frac{2\pi (E/e)\alpha_p\sigma_p^2\sigma_z}{G_{ic} \frac{\text{Re}\left[Z_\parallel(k_c)\right]}{k_c}},
    \label{eq:Ith-emprical}
\end{equation}
where we define $G_{ic}=G_i(x(k_c)) \approx 1.75$. The latter value follows from solving Eq.~\eqref{eq:GiGr_vs_ZrZi} in the limit of $\left . \text{Re}\left[Z_\parallel(k)\right]/\text{Im}\left[Z_\parallel(k)\right]\right|_{k=k_c}\rightarrow\infty$~\cite{blednykh2023microwave}. The properties of the impedance ensure that $\text{Re}\left[Z_\parallel(k)\right]>0$ when $k\neq 0$. Equation~\eqref{eq:Ith-emprical} highlights the importance of the real part of the impedance in determining the threshold current. However, the chamber shielding fundamentally determines $k_c$, which means that the imaginary part of the impedance also plays a significant role in the overall result. An alternative approach, as demonstrated in~\cite{stupakov2002beam}, is to use the cutoff wavenumber of the vacuum chamber (whether straight or curved) for $k_c$, where $\text{Im}[Z_\parallel(k_c)] \approx 0$, to determine the minimum threshold current.

By connecting Eqs.~\eqref{eq:boussard-criterion},~\eqref{eq:Ith-general} and~\eqref{eq:Ith-emprical}, the effective impedance $\left| \frac{Z}{n} \right|$ for high-frequency broadband impedances can be evaluated as
\begin{align}
    \left| \frac{Z}{n} \right|
    & = \frac{\sqrt{2\pi}}{C} \text{Min} \left[ \frac{G_i(k) \left| Z_\parallel(k) \right|^2}{k\text{Re}\left[Z_\parallel(k)\right]} \right] \nonumber \\
    & \approx \frac{\sqrt{2\pi} G_{ic} \text{Re}\left[Z_\parallel(k_c)\right]}{Ck_c}.
\end{align}
The above equations offer a practical method for converting the calculated impedances into an effective impedance, enabling the estimation of the threshold current for instability.

By applying the steady-state CSR impedance model with parallel-plates shielding~\cite{agoh2004calculation} to Eq.~\eqref{eq:Ith-general} and taking the minimum value, the CSR-induced instability threshold current is obtained as ~\cite{cai2011theory, dastan2024coherent}:
\begin{equation}
    I_{th}(\text{CSR})
    \approx \frac{6\sqrt{2}(E/e)\alpha_p\sigma_p^2\sigma_z}{\sqrt{\pi}Z_0h},
    \label{eq:Ith-csr-ppss}
\end{equation}
where $Z_0$ is the impedance of the free space and $2h$ is the full height of the vacuum chamber in bending magnets. The critical wavenumber for Eq.~\eqref{eq:Ith-csr-ppss} is
\begin{equation}
    k_c\approx 2\sqrt{\rho/h^3}
    \label{eq:kc-CSR}
\end{equation}
with $\rho$ the bending radius in the bending magnets. Comparing Eqs.~\eqref{eq:Ith-csr-ppss} and~\eqref{eq:boussard-criterion}, one can find that the CSR impedance with parallel-plates shielding contributes to an effective impedance of
\begin{equation}
    \left| \frac{Z}{n} \right|_\text{CSR}
    \approx \frac{\pi^2h Z_0}{3C}.
    \label{eq:Zeff-CSR}
\end{equation}
Comparing with Eq.~(2) in~\cite{venturini2005coherent}, the above expression (with $C$ replaced by $2\pi R$) represents a value close to the peak of $\text{Re}Z(n)/n$ defined therein.

A similar analysis can be performed using the CWR impedance to estimate the instability threshold, as detailed in~\cite{blednykh2023microwave}. In Sec.~\ref{sec:cwi}, we will extend this analysis by explicitly considering the impact of storage ring parameters in the presence of wigglers.



\section{Analytic models for CWR impedance}\label{sec:cwr-impedance}


The CWR impedance has been analyzed extensively~\cite{wu2003calculation, stupakov2016analytical}, with particular emphasis on the low-frequency regime relevant to microwave instability~\cite{wu2003impact, blednykh2023microwave}. In what follows, we revisit these analytic treatments in a unified framework, contrasting free-space and parallel-plate shielding, and distinguishing steady-state emission in infinitely long wigglers from transient effects in finite-length devices. Although free-space models are not adequate for quantitative predictions of CWR effects, we retain them as a compact analytic baseline that clarifies how shielding alters the impedance and helps interpret instability mechanisms. This organization provides a coherent reference against which the subsequent, more realistic formulations can be assessed for scope and applicability.

The simplest broadband model for the CWR impedance is the free-space steady-state model (hereafter abbreviated as FS-SS model) per unit length~\cite{wu2003calculation, stupakov2016analytical}:
\begin{equation}
\frac{Z_\parallel(k)}{L}\approx
\frac{Z_0k_wk}{4k_0}
\left[
1 - \frac{2i}{\pi}\ln\left( \frac{k}{k_0} \right)
\right],
\label{eq:CWR_Impedance_FS-SS}
\end{equation}
where $k_w=2\pi/\lambda_w$ is the wiggler wavenumber, with $\lambda_w$ being the wiggler period length, and $k_0=4k_w/\theta_0^2$ is the fundamental radiation wavenumber of the wiggler. The parameter $\theta_0=K/\gamma$ represents the deflection angle, where $K\approx 93.4B_w\lambda_w$ is the wiggler strength ($B_w$ is the peak field of the wiggler in Tesla) and $\gamma$ is the Lorentz factor. Equation~\eqref{eq:CWR_Impedance_FS-SS} is valid under the conditions $r = k/k_0\ll 1 $, $\theta_0 \ll 1$ , and $K\gg 1$~\cite{wu2003calculation}. Under the same conditions, analytical models have been developed to account for the shielding of parallel plates~\cite{stupakov2016analytical, blednykh2023microwave}.

We begin with Eq.~(C7) from~\cite{stupakov2016analytical} to derive simplified models of the CWR impedance, specifically tailored for wigglers of finite length. In doing so, we incorporate parallel-plates shielding along with transient effects at both the entrance and exit of the wigglers (hereafter abbreviated as PP-TR model). For simplicity, we retain only the terms corresponding to $n=0,\pm 1$. The resulting impedance per unit length of the wiggler is then expressed as:
\begin{equation}
    \frac{Z_\parallel(k)}{L}\approx
    \frac{(i-1)Z_0\theta_0^2k}{2\sqrt{\pi}}
    \sum_{p=0}^\infty F_p(k,u),
    \label{eq:CWR_Impedance_PP-TR}
\end{equation}
where $F_p(k,u)$ is given by:
\begin{equation}
    F_p(k,u)=
    \frac{1}{\sqrt{kk_w}a} \int_0^u 
    \frac{d\nu}{\sqrt{\nu}}
    \sin^2\frac{\nu}{2} e^{-i\nu \beta_p}
    G(\nu,r,u),
    \label{eq:Fp1}
\end{equation}
and $G(\nu,r,u)$ is defined as:
\begin{align}
    G(\nu,r,u)=
    & e^{i\frac{4r}{\nu}\sin^2 \frac{\nu}{2}}
    \left[ 
    \left( 1-\frac{\nu+\sin\nu}{u} \right) J_0(R) \right. \nonumber \\
    & \left. + i \left( 1-\frac{\nu+2\sin\nu+\frac{1}{2}\sin(2\nu)}{u} \right)
    J_1(R)
    \right].
    \label{eq:G1}
\end{align}
Here, $J_n(x)$ is the Bessel function of the first kind, $a$ denotes the full distance between the parallel plates, and $u=k_wL_w=2\pi N_p$, where $L_w$ is the total length of the wiggler and $N_p$ is the number of periods. The coefficient $\beta_p$ in the phasor is given by:
\begin{equation}
    \beta_p = \frac{k_p^2}{2k k_w} + r,
    \label{eq:CWR_modulation}
\end{equation}
with $k_p=(2p+1)\pi/a$ for $p=0,1,2,\dots$. Finally, the parameter $R$ is defined as
\begin{equation}
    R = r \left( \sin\nu - \frac{4}{\nu} \sin^2 \frac{\nu}{2} \right).
    \label{eq:R1}
\end{equation}

The first term on the right-hand side of Eq.~\eqref{eq:CWR_modulation} represents shielding effects, while the second reflects the influence of the beam's wiggling motion. The shielding effects are more pronounced in the low-frequency region, whereas the wiggling-motion effects dominate in the high-frequency region. In physical space, these correspond to influencing the long-range and short-range wakefields, respectively.

The terms proportional to $1/u$ in $G(\nu,r,u)$ represent the transient effects in the CWR. In the limit of $u \rightarrow \infty$, the expression simplifies to:
\begin{equation}
    G(\nu,r,u)|_{u \rightarrow \infty}=
    e^{i\frac{4r}{\nu}\sin^2 \frac{\nu}{2}}
    \left[ 
    J_0(R)
    + i
    J_1(R)
    \right].
    \label{eq:G2}
\end{equation}
This result corresponds to Eq.~(D4) in~\cite{stupakov2016analytical}, which describes the steady-state CWR impedance for an infinitely long wiggler with parallel-plates shielding (hereafter abbreviated as the PP-SS model).

For the CWR in free space, one can take the limit of $a \rightarrow \infty$ in Eq.~\eqref{eq:CWR_Impedance_PP-TR}. Then, the summation over $p$ is replaced by integration~\cite{stupakov2016analytical}, leading to
\begin{align}
    F(k,u) & =
    \left. \sum_{p=0}^\infty F_p(k,u) \right|_{a \rightarrow \infty} \nonumber \\
    & = \frac{1-i}{4\sqrt{\pi}}
    \int_0^u \frac{d\nu}{\nu}
    \sin^2\frac{\nu}{2} e^{-i\nu r}
    G(\nu,r,u).
    \label{eq:F1}
\end{align}
Substituting this expression into Eq.~\eqref{eq:CWR_Impedance_PP-TR} yields the free-space CWR impedance model that includes transient effects (hereafter referred to as the FS-TR model).

We next examine two regimes where further simplifications can provide deeper insights into the nature of CWR: the low-frequency regime ($ r \ll 1 $) and the high-frequency regime ($ r \gtrsim 1 $).

\subsection{Low-frequency regime $ r \ll 1 $}

For small $r$, the Bessel functions in Eq.~\eqref{eq:G1} can be approximated as $J_0(R) \approx 1$ and $J_1(R) \approx 0$. Under this approximation, only the leading terms of the slowly modulating phasors are retained. As a result, Eq.~\eqref{eq:Fp1} simplifies to:
\begin{equation}
    F_p(k,u)=
    \frac{1}{\sqrt{kk_w}a} \int_0^u 
    \frac{d\nu}{\sqrt{\nu}}
    \sin^2\frac{\nu}{2} e^{-i\nu \beta_p}
    \left( 1-\frac{\nu+\sin\nu}{u} \right).
    \label{eq:Fp2}
\end{equation}
The integral over $\nu$ in Eq.~\eqref{eq:Fp2} can be evaluated analytically, resulting in the following expression:
\begin{equation}
    F_p(k,u)=F_{p1}(k,u) + F_{p2}(k,u) + F_{p3}(k,u) + F_{p4}(k,u),
    \label{eq:Fp3}
\end{equation}
where the individual terms are given by:
\begin{equation}
    F_{p1}(k,u)=
    -\frac{\sqrt{\pi u}}{4\sqrt{kk_w}a} 
    \left[ E_1\left( \phi_{-1} \right) - 2 E_1\left( \phi_{0} \right) + E_1\left( \phi_{1} \right) \right],
\end{equation}
\begin{equation}
    F_{p2}(k,u)=
    -\frac{\sqrt{u}}{4\sqrt{kk_w}a} 
    \left[ E_2\left( \phi_{-1}^2 \right) - 2 E_2\left( \phi_{0}^2 \right) + E_2\left( \phi_{1}^2 \right) \right],
\end{equation}
\begin{equation}
    F_{p3}(k,u)=
    \frac{\sqrt{\pi u}}{8\sqrt{kk_w}a} 
    \left[ E_3\left( \phi_{-1} \right) - 2 E_3\left( \phi_{0} \right) + E_3\left( \phi_{1} \right) \right],
\end{equation}
\begin{align}
    F_{p4}(k,u)=
    & -\frac{i\sqrt{\pi}}{8\sqrt{kk_wu}a} 
    \left[ E_1\left( \phi_{-2} \right) - 2 E_1\left( \phi_{-1} \right) \right.\nonumber \\
    & \left. + 2E_1\left( \phi_{1} \right) - E_1\left( \phi_{2} \right) \right].
\end{align}
Here, the parameter $\phi_{n}$ is defined as $\phi_{n}=\sqrt{i (\beta_p+n) u}$, and the functions $E_1(x)$, $E_2(x)$, and $E_3(x)$ are given by:
\begin{equation}
    E_1(x)=\frac{\text{erf}(x)}{x}, \ 
    E_2(x)=\frac{e^{-x}}{x}, \
    E_3(x)=\frac{\text{erf}(x)}{x^3},
\end{equation}
where $\text{erf}(x)$ is the error function, defined as $\text{erf}(x)=\frac{2}{\sqrt{\pi}}\int_0^x e^{-t^2} dt$.

As $u \rightarrow \infty$, $F_{pj} \rightarrow 0$ for $j=2,3,4$ and $\text{erf}(x) \rightarrow 1$ as $|x| \rightarrow \infty$. Consequently $F_p$ simplifies to the following:
\begin{align}
    F_p(k,\infty) & = F_{p1}(k,\infty) \nonumber \\
    & =
    -\frac{\sqrt{\pi}}{4\sqrt{kk_w}a}
    \left[ \frac{1}{\sqrt{i\beta_p^-}} - \frac{2}{\sqrt{i\beta_p}}
    + \frac{1}{\sqrt{i\beta_p^+}} \right],
    \label{eq:Fp4}
\end{align}
with $\beta_p^{\pm}=\beta_p\pm 1$. This result is equivalent to Eq.~(11) of~\cite{blednykh2023microwave}, but presented in a more concise form.

The terms involving $\phi_{\pm 2}$ in $F_{p4}$ originate from the wiggler's second harmonic and are therefore negligible. Since $u \gg 1$ under typical conditions, particularly for long wigglers where $N_p \gg 1$, the magnitude of $F_{p4}$ is on the order of $1/u$ relative to $F_{p1}$. Consequently, $F_{p4}$ makes an insignificant contribution to $F_p$.

\subsection{High-frequency regime $r \gtrsim 1$}

In the high-frequency regime, where $r$ is relatively large, short-range CWR wakefields dominate the total impedance. In this scenario, only the terms in $G(\nu,r,u)$ with $\nu \ll 1$ contribute significantly to $F_p(k,u)$, while transient effects become negligible. Taking these factors into account, Eq.~\eqref{eq:Fp1} can be approximated as:
\begin{align}
    F_p(k,\infty) \approx
    & \frac{1}{4\sqrt{kk_w}a} \int_0^\infty d\nu
    \nu^{3/2} e^{-i\nu \frac{k_p^2}{2kk_w}}
    e^{-i\frac{r\nu^3}{12}} \nonumber \\
    & \times \left[ J_0\left(\frac{r\nu^3}{12}\right) - iJ_1\left(\frac{r\nu^3}{12}\right) \right]
    .
    \label{eq:Fp5}
\end{align}
The integral over $\nu$ converges poorly in the above equation. As an alternative, we start from Eq.~(C2) in~\cite{stupakov2016analytical} and retain the integration over $\xi$. By Taylor expanding the $\nu$-dependent terms around 0 and keeping only the leading-order terms, we obtain
\begin{align}
    F_p(k,\infty)\approx
    & \frac{1}{4\pi\sqrt{kk_w}a}
    \int_0^{2\pi} d\xi \left|\cos\xi\right|^{2/3} \nonumber \\
    & \int_0^\infty d\nu' \nu'^{3/2} e^{-i\nu' \frac{k_p^2}{2kk_w}}
    e^{-i\frac{r\nu'^3}{6}},
    \label{eq:Fp6}
\end{align}
where a change of variable $\nu'=\nu\cos^{2/3}\xi$ has been used. To simplify the expression, the variable $\nu$ within the factor $\sqrt{\nu}\exp\left( -i\nu k_p^2/(2kk_w) \right)$ is directly replaced by $\nu'$, neglecting the implicit dependence on $\xi$. It is worth noting that the integral over $\xi$ in Eq.~\eqref{eq:Fp6} matches the one in Eq.~(14) of~\cite{wu2003calculation}. In the limit $a\rightarrow \infty$, the discrete summation over $p$ becomes a continuous integration, and Eq.~\eqref{eq:Fp6} reduces to Eq.~(24) of~\cite{wu2003calculation}. Evaluating the integrals over $\xi$ and $\nu'$, the final expression for $F_p(k,\infty)$ can be expressed in terms of Airy functions and their derivatives:
\begin{align}
    F_p(k,\infty)\approx
    & -\frac{\pi(-1)^{7/12}\Gamma\left(\frac{5}{6}\right)}{\Gamma\left(\frac{4}{3}\right) \sqrt{kk_w}a(2r)^{5/6}} \nonumber \\
    & \times \left[
    \Pi_p \left[ \text{Ai}^2(\Pi_p)+\text{Bi}^2(\Pi_p) \right] \right. \nonumber \\
    & \left. + \text{Ai}'^2(\Pi_p)+\text{Bi}'^2(\Pi_p)
    \right],
    \label{eq:Fp7}
\end{align}
where $\Pi_p$ is given by:
\begin{equation}
    \Pi_p=-\frac{(-1)^{1/3}k_p^2}{2kk_w(2r)^{1/3}}.
\end{equation}

In the free-space scenario, Eq.~\eqref{eq:F1} is approximated as:
\begin{align}
    F(k,\infty) \approx
    & \frac{1-i}{16\sqrt{\pi}}
    \int_0^\infty d\nu
    \nu e^{-i\frac{r\nu^3}{12}} \nonumber \\
    & \times \left[ J_0\left(\frac{r\nu^3}{12}\right) - iJ_1\left(\frac{r\nu^3}{12}\right) \right].
\end{align}
Performing the integration over $\nu$ yields:
\begin{equation}
    F(k,\infty) \approx
    \frac{(1-i) 6^{2/3}\Gamma\left( \frac{2}{3} \right)}{4 (ir)^{2/3} \Gamma\left( \frac{1}{3} \right) \Gamma\left( \frac{1}{6} \right)}.
    \label{eq:F2}
\end{equation}
Substituting Eq.~\eqref{eq:F2} into Eq.~\eqref{eq:CWR_Impedance_PP-TR}, we obtain the high-frequency steady-state CWR impedance in free space as
\begin{equation}
    \frac{Z_\parallel(k)}{L}=
    \frac{(\sqrt{3}+i) 3^{2/3} \Gamma\left( \frac{2}{3} \right)}{2\sqrt{\pi} \Gamma\left( \frac{1}{3} \right) \Gamma\left( \frac{1}{6} \right)}
    Z_0 \left( k_w\theta_0 \right)^{2/3} k^{1/3}.
    \label{eq:CWR_Impedance_FS-SS_large_k}
\end{equation}
This result exactly reproduces Eq.~(27) in~\cite{wu2003calculation}. In particular, Eq.~\eqref{eq:CWR_Impedance_FS-SS_large_k} exhibits the same scaling law with respect to $k$ as the high-frequency limit of the steady-state CSR impedance in free space~\cite{faltens1973longitudinal}. This similarity is expected, considering that short-range CSR fields are influenced by local curvature of the beam orbit.

Referring to Eq.~(1) in~\cite{dastan2024coherent}, there is a straightforward way to derive the CWR impedance of Eq.~\eqref{eq:CWR_Impedance_FS-SS_large_k}. Assuming that the curvature, expressed as $\rho^{-1}$, depends on $z$ as
\begin{equation}
    \rho^{-1}(z)=k_w\theta_0 |\cos(k_wz)|,
\end{equation}
one can take the average of $\rho^{-2/3}$ over $z$, resulting in:
\begin{equation}
    \langle \rho^{-2/3} \rangle =
    \frac{\Gamma \left( \frac{5}{6} \right)}{\sqrt{\pi} \Gamma \left( \frac{4}{3} \right) } \left(k_w\theta_0\right)^{2/3}.
\end{equation}
Substituting this result into Eq.~(1) in~\cite{dastan2024coherent} reproduces Eq.~(27) of~\cite{wu2003calculation} exactly.

It should be emphasized that the total CWR impedances, computed using Eq.~\eqref{eq:Fp7} for parallel-plates shielding and Eq.~\eqref{eq:CWR_Impedance_FS-SS_large_k} in the free space for the regime $r \gg 1$, depend solely on the peak magnetic field and total length of the wiggler, and are independent of the wiggler period $\lambda_w$. However, in the regime $r \lesssim 1$, the wiggler period $\lambda_w$ has a significant impact on the scaling behavior of the CWR impedances, as will be demonstrated in Fig.~\ref{fig:FS-PP-comparison}.


\section{Impact of CWR impedance in Tau-Charm factories}\label{sec:cwi}

By substituting the CWR impedance from Eq.~\eqref{eq:CWR_Impedance_FS-SS} into Eq.~\eqref{eq:Ith-general}, the threshold current as a function of wavenumber can be approximated as~\cite{blednykh2023microwave}
\begin{equation}
    I_{th}(k) \approx \frac{8\pi \sqrt{2\pi}(E/e)\alpha_p\sigma_p^2\sigma_z}{L_wZ_0\theta_0^2 \ln \frac{k_0}{k}}.
    \label{eq:Ith_cwr_simple}
\end{equation}
The validity of the above expression is subject to the conditions $1/\sigma_z \ll k \ll k_0$. The lower bound arises from the assumptions underlying the instability analysis~\cite{stupakov2002beam}, while the upper bound reflects the applicability range of the impedance model used.

When considering the shielding effects of parallel plates, the minimum threshold current due to CWR impedance can be approximated at the cutoff wavenumber
\begin{equation}
    k_c(\text{CWR})=\frac{k_w^2+\frac{\pi^2}{a^2}}{2k_w},
    \label{eq:kc-CWR}
\end{equation}
leading to
\begin{equation}
I_{th}(\text{CWR})\approx
\frac{8\pi \sqrt{2\pi}(E/e)\alpha_p\sigma_p^2\sigma_z}{L_wZ_0\theta_0^2 \ln \frac{k_0}{k_c}}.
\label{eq:Ith_cwr_kc}
\end{equation}
As pointed out in~\cite{emma2001systematic, wu2003impact}, the beam parameters $\alpha_p$, $\sigma_y$ and $\sigma_z$ are not independent but are instead strongly correlated through the properties of the ring lattice. The momentum compaction is defined as $\alpha_p=I_1/C$ with $I_1=\int_0^C \frac{\eta_x}{\rho}ds$ the first synchrotron integral and $\eta_x$ the horizontal dispersion function. The wiggler contribution to $I_1$ can be approximated as $I_{1w} \approx -L_w/(2\rho_w^2 k_w^2)$~\cite{wolski2003accelerator}, with $\rho_w = E/(e c B_w)$. For Tau-Charm factories such as STCF, this contribution is very small, so $\alpha_p$ is essentially determined by the arc optics, with damping wigglers having negligible influence. The energy spread is determined by
\begin{equation}
    \sigma_p^2 = C_q \gamma^2 \frac{I_3}{j_zI_2},
\end{equation}
where $C_q=3.8319\times 10^{-13}$ m is the quantum radiation constant; $j_z$ is the longitudinal damping partition number; the second and third synchrotron integrals are given by $I_2=\int_0^C \frac{1}{\rho^2}ds$ and $I_3=\int_0^C \frac{1}{|\rho|^3}ds$. The contributions from arc dipoles and wigglers to these integrals can be explicitly formulated as follows:
\begin{equation}
    I_{2a}=\frac{2\pi B_a}{B\rho}, \quad I_{2w} = \frac{L_wB_w^2}{2(B\rho)^2},
\end{equation}
\begin{equation}
    I_{3a}=\frac{2\pi B_a^2}{(B\rho)^2}, \quad I_{3w} =\frac{4L_wB_w^3}{3\pi(B\rho)^3},
\end{equation}
where $B\rho \approx E/(ec)$ represents the beam rigidity. Here, it is assumed that there is only one type of arc bending magnet, characterized by a uniform magnetic field $B_a$, and that the wigglers have sinusoidal field distributions with a peak magnetic field $B_w$. Consequently, the energy spread is expressed as
\begin{equation}
    \sigma_p^2 =
    \frac{C_q\gamma^2\left( 3 \pi B_a + 8B_wF_w \right)}{3\pi j_z (B\rho) \left( 1 + F_w \right)},
\end{equation}
with $F_w=I_{2w}/I_{2a}$ following the definition in~\cite{wu2003impact}. Using the bunch length defined as $\sigma_z=\alpha_pC\sigma_p/(2\pi \nu_s)$ and the synchrotron tune given by
\begin{equation}
    \nu_s=\frac{1}{2\pi} \sqrt{-\frac{\alpha_p V_\text{rf}\omega_\text{rf}C\cos \phi_s}{c(E/e)}},
\end{equation}
the scaling law for the threshold current driven by CWR can be derived as
\begin{equation}
    I_{th}(\text{CWR})\propto
    \frac{\gamma^3\alpha_p^{3/2}\sqrt{C}}{\sqrt{V_\text{rf}}L_w\theta_0^2 \ln \frac{k_0}{k_c}}
    \left( \frac{3\pi B_a+8B_wF_w}{1 + F_w} \right)^{3/2}.
    \label{eq:Ith_cwr_scaling_law}
\end{equation}
It should be noted that this formulation differs slightly from Eq.~(24) in~\cite{wu2003impact} and is better suited for analyzing instability thresholds in Tau-Charm factory rings. For a given beam energy and storage-ring layout, the parameter $F_w$ is primarily determined by the desired vertical damping time:
\begin{equation}
    \tau_y =
    \frac{3C}{\gamma^3r_ecI_2}
    =\frac{3Cm_ec^2}{2\pi \gamma^2 r_eec^2 B_a(1+F_w)}.
    \label{eq:DampingTime}
\end{equation}
While increasing the momentum compaction is a straightforward approach to raising the instability threshold, it generally requires larger dispersion functions in the arcs, which consequently increases the horizontal emittance. If we assume the horizontal emittance to be fixed, thus preventing any further increase in $\alpha_p$, Eq.~\eqref{eq:Ith_cwr_scaling_law} indicates that the remaining viable methods to improve the threshold current are reducing the wiggler period length $\lambda_w$, as will be demonstrated using the STCF project as an example in Sec.~\ref{sec:simulations}, and decreasing the wiggler chamber height $a$.

We apply Eqs.~\eqref{eq:Ith-csr-ppss} and~\eqref{eq:Ith_cwr_kc} to STCF using beam and wiggler parameters as summarized in Tab.~\ref{tab:stcf} (for detailed design configurations of STCF, see~\cite{zou2025eefact, zhang2025eefact, zou2025opticsdesignsupertaucharm}). The parameters shared by designs for different beam energies are: the ring circumference $C$=860.321 m, number of bunches $N_b$=688, full chamber height in the bending magnets $2h$=52 mm, full chamber height in the wigglers $a$=42 mm, number of wigglers $N_w$=16, number of periods per wiggler $N_p$=6, wiggler's period length $\lambda_w$=0.8 m, wiggler's peak field $B_w$=1.6 T. The threshold currents due to CSR and CWR are compared with the design bunch currents to achieve the relevant luminosity goals. It is seen that, at the lowest beam energy of 1 GeV, the current STCF design might operate below the instability threshold, setting an obstacle to achieving the luminosity goal.
\begin{table}
    \centering
    \begin{tabular}{cccccc}
    \toprule
        $E$ & GeV & 1 & 1.5 & 2 & 3.5 \\
        $\alpha_p$ & $10^{-3}$ & 1.26 & 1.32 & 1.35 & 1.37 \\
        $V_\text{rf}$ & MV & 0.75 & 1.2 & 2.5 & 6 \\
        $U_0$ & keV & 106 & 267 & 543 & 1494 \\
        $\nu_s$ & - & 0.0146 & 0.0154 & 0.0194 & 0.0228 \\
        $\sigma_z$ & mm & 6.62 & 7.89 & 7.21 & 8.26 \\
        $\sigma_p$ & $10^{-4}$ & 6.18 & 6.93 & 7.8 & 10.02 \\
        $\tau_z$ & ms & 27 & 16 & 11 & 6.7 \\ 
        $F_w$ & - & 12.51 & 5.56 & 3.13 & 0.0 \\    
        $k_0$ & mm$^{-1}$ & 8.42 & 18.94 & 33.67 & 103.12 \\ 
        $k_c$ & mm$^{-1}$ & 0.36 & 0.36 & 0.36 & 0.36 \\ 
        $I_b$(Design) & mA & 1.6 & 2.47 & 2.91 & 2.91 \\ 
        $I_{th}(\text{CSR})$ & mA & 1.56 & 3.67 & 5.79 & 19.44 \\
        $I_{th}(\text{CWR})$ & mA & 0.59 & 2.49 & 6.09 & 50.24 \\
    \hline
    \end{tabular}
    \caption{Beam and wiggler parameters optimized at different beam energies, and derived parameters relevant to CSR and CWR for STCF project. The threshold bunch current $I_{th}(\text{CSR})$ and $I_{th}(\text{CWR})$ are calculated by Eqs.~\eqref{eq:Ith-csr-ppss} and~\eqref{eq:Ith_cwr_simple}, respectively.}
    \label{tab:stcf}
\end{table}

The threshold of Eq.~\eqref{eq:Ith_cwr_kc} serves a fast check of CWR effects for storage rings, but is applicable only when $1/\sigma_z \ll k_c(\text{CWR}) \ll k_0$. Otherwise, different approaches have to be found. For example, for $k \gg k_0$, the impedance of Eq.~\eqref{eq:CWR_Impedance_FS-SS_large_k} can be applied to Eq.~\eqref{eq:Ith-general} to determine the instability threshold. Furthermore, with detailed configurations of the rings, one can use the calculated impedance and beam parameters to perform numerical simulations to determine the threshold, as will be discussed in the following section.

\section{Simulations of CWR-driven microwave instability for STCF project}\label{sec:simulations}

As established in the previous section, CSR and CWR impedances exert their strongest influence at the lowest beam energy (1 GeV) of the STCF. Thus, this section specifically focuses on this scenario. We begin by comparing the various CWR impedance models introduced in Sec.~\ref{sec:cwr-impedance}, utilizing the beam and wiggler parameters detailed in Table~\ref{tab:stcf}.
\begin{figure}[htbp]
  \centering
    \includegraphics[width=\linewidth]{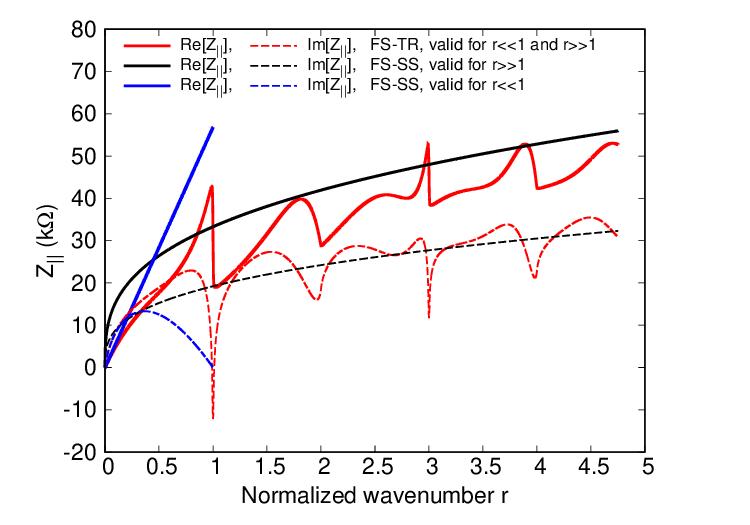}
  \caption{Total CWR impedance of the STCF rings calculated with free-space models. Beam and wiggler parameters correspond to the 1 GeV case (see Table~\ref{tab:stcf} and related discussion). The red curves are obtained from Eq.~\eqref{eq:CWR_Impedance_PP-TR} together with Eq.~\eqref{eq:F1}, while the blue and black curves are calculated from Eq.~\eqref{eq:CWR_Impedance_FS-SS} and Eq.~\eqref{eq:CWR_Impedance_FS-SS_large_k}, respectively.
  }
  \label{fig:ZL-FS-comparison}
\end{figure}
\begin{figure}[htbp]
  \centering
    \includegraphics[width=\linewidth]{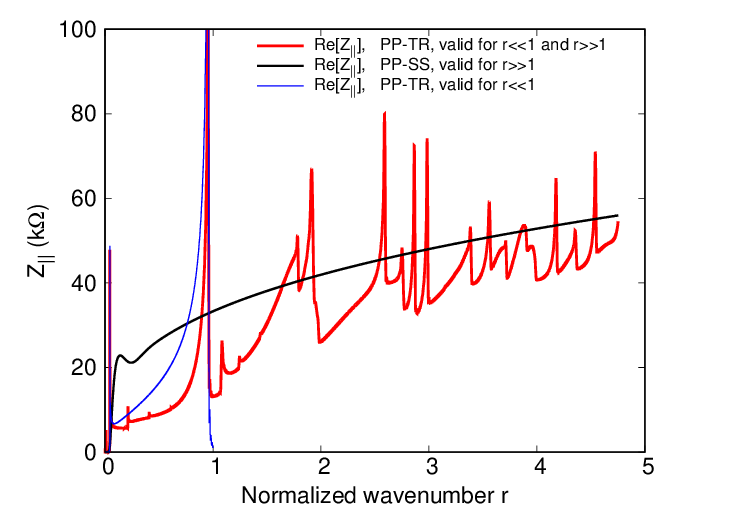}
    \includegraphics[width=\linewidth]{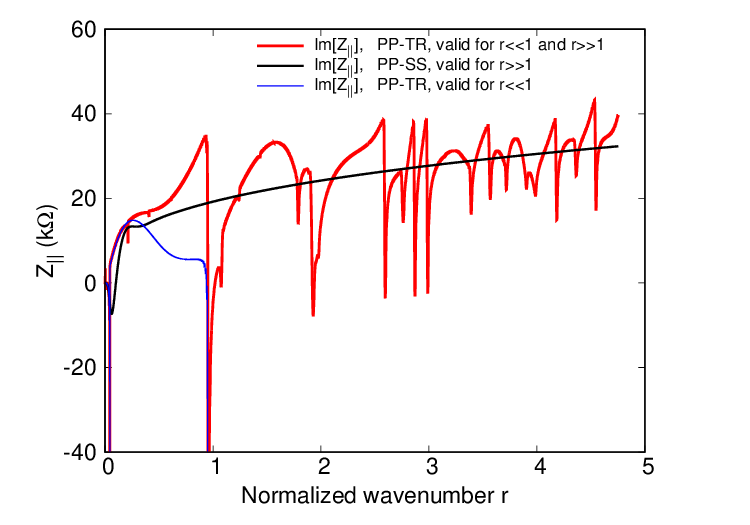}
  \caption{Total CWR impedance of the STCF rings calculated with parallel-plates shielding models. Beam and wiggler parameters correspond to the 1 GeV case (see Table~\ref{tab:stcf} and related discussion). The red curves are computed using Eq.~\eqref{eq:CWR_Impedance_PP-TR} with Eq.~\eqref{eq:Fp1}, while the blue and black curves corresponds to calculations using Eq.~\eqref{eq:Fp3} and Eq.~\eqref{eq:Fp7}, respectively.
  }
  \label{fig:ZL-PP-comparison}
\end{figure}

We first compare the steady-state and transient free-space models for calculating the total CWR impedance of the STCF rings, as illustrated in Fig.~\ref{fig:ZL-FS-comparison}. Notably, the FS-TR model described by Eq.~\eqref{eq:F1} clearly differs from the formulations given in~\cite{wu2003calculation}, yet successfully reproduces the key features of CWR impedance shown in Figs. 4 and 5 of that reference. The FS-SS model defined by Eq.~\eqref{eq:CWR_Impedance_FS-SS_large_k} correctly captures the scaling behavior of CWR impedance in the high-frequency regime but deviates significantly from the other two models at low frequencies. As a result, the FS-SS model of Eq.~\eqref{eq:CWR_Impedance_FS-SS_large_k} is not suitable for simulations where low-frequency components dominate the instability threshold, as is the case for STCF, where the bunch length is approximately 7 mm.

Figure~\ref{fig:ZL-PP-comparison} shows the total CWR impedance of the STCF rings calculated using three parallel-plates shielding models. The presence of parallel plates alters the resonance conditions~\cite{blednykh2023microwave}, resulting in additional resonance spikes in the CWR impedance compared to the free-space models discussed earlier. Nevertheless, in the high-frequency regime, the overall impedance amplitude continues to follow the expected scaling law, $Z_\parallel(k) \propto k^{1/3}$.
\begin{figure*}
  \centering
    \includegraphics[width=0.9\linewidth]{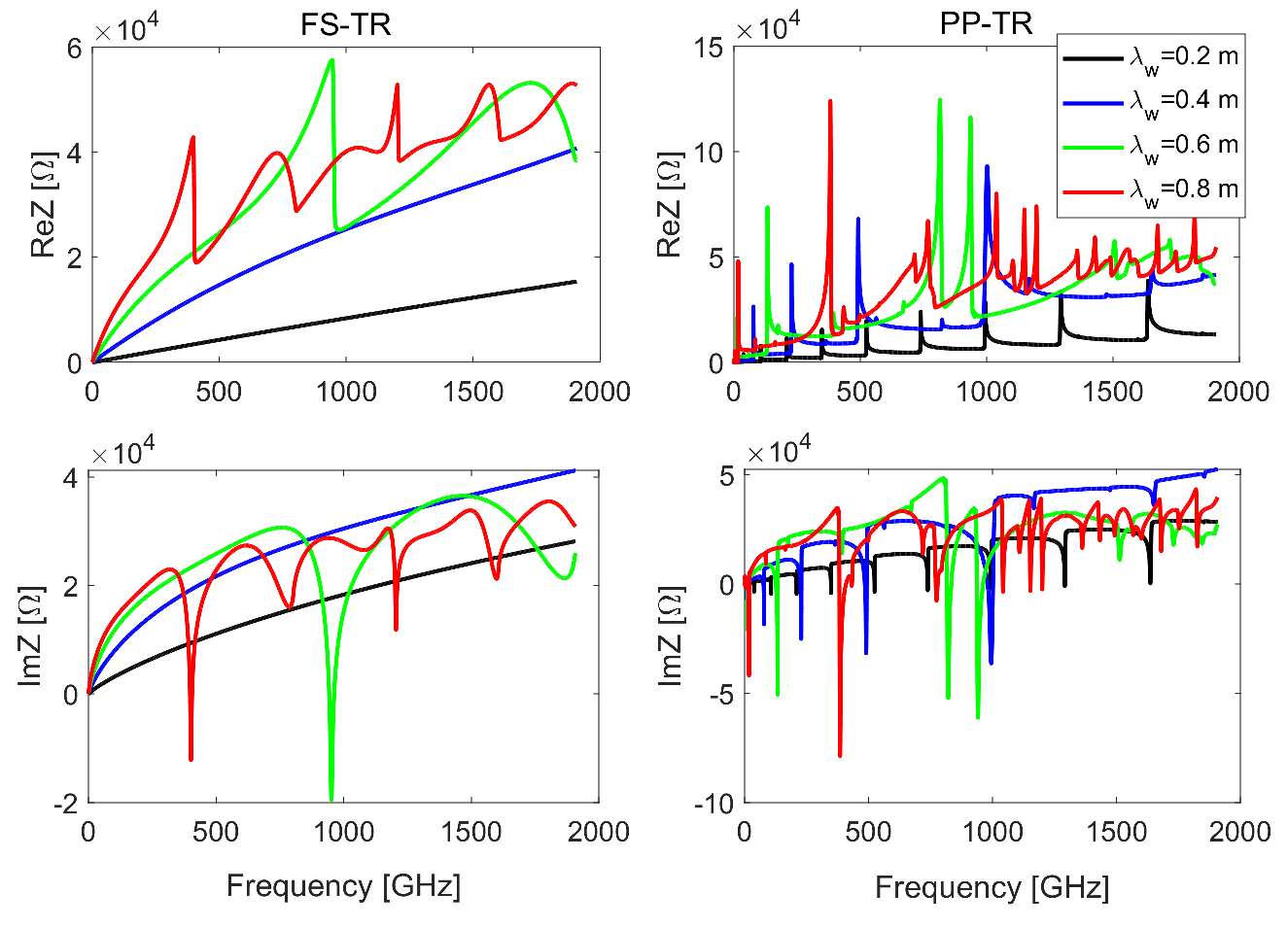}
  \caption{Total CWR impedance of the STCF rings for different wiggler periods, calculated using free-space (left) and parallel-plate shielding (right) models. The total wiggler length and peak field are kept fixed while varying the period length. The legend gives the wiggler period length.
  }
  \label{fig:FS-PP-comparison}
\end{figure*}

It is important to emphasize that the CWR impedance models given by Eq.~\eqref{eq:CWR_Impedance_FS-SS} and Eq.~\eqref{eq:Fp3} are valid only in the regime where $k\ll k_0$. When $k\gtrsim k_0$, these models exhibit noticeable deviations from the more accurate transient models, as illustrated in Figs.~\ref{fig:ZL-FS-comparison} and~\ref{fig:ZL-PP-comparison}. Their applicability must be carefully assessed before being used in instability analyses or numerical simulations. Accordingly, in what follows we employ the FS-TR and PP-TR models to simulate CWR effects in STCF at 1 GeV. The FS-TR model serves as a compact analytic baseline for interpreting the impact of chamber shielding and for instability estimates, while the PP-TR model is used for quantitative predictions.

We employ the STABLE code to perform tracking simulations to evaluate CWR effects in STCF. The STABLE code~\cite{HeGPU2021}, originally developed as a GPU-accelerated program for multi-bunch longitudinal beam dynamics simulations, has been adapted with minor modifications for single-bunch longitudinal beam dynamics tracking. It efficiently supports simulations involving tens of millions of macro-particles and tens of thousands of slices while maintaining high computational efficiency. In the following analysis, we apply both the FS-TR and PP-TR CWR impedance models in the tracking simulations for STCF.

\begin{figure*}[htbp]
  \centering
    \includegraphics[width=0.49\textwidth]{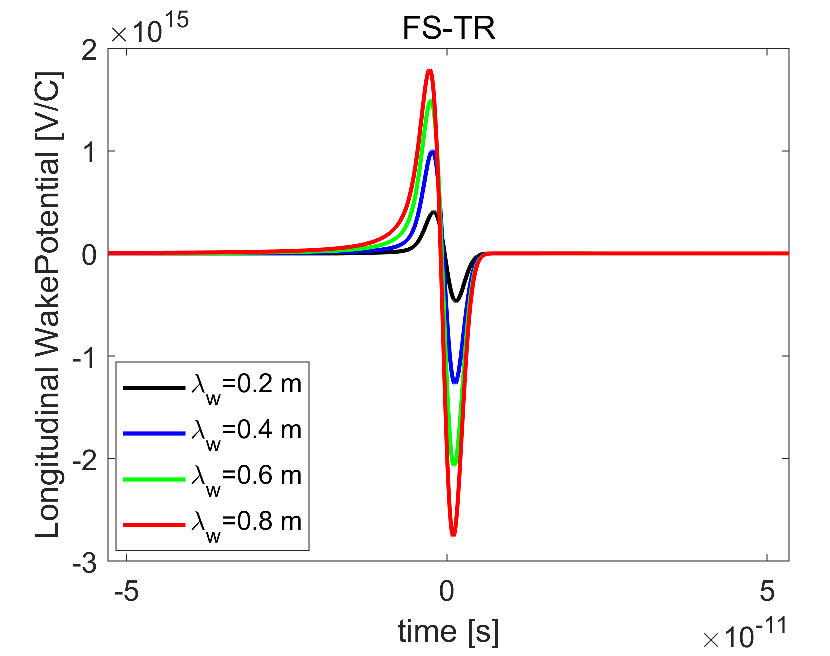}
    \includegraphics[width=0.49\textwidth]{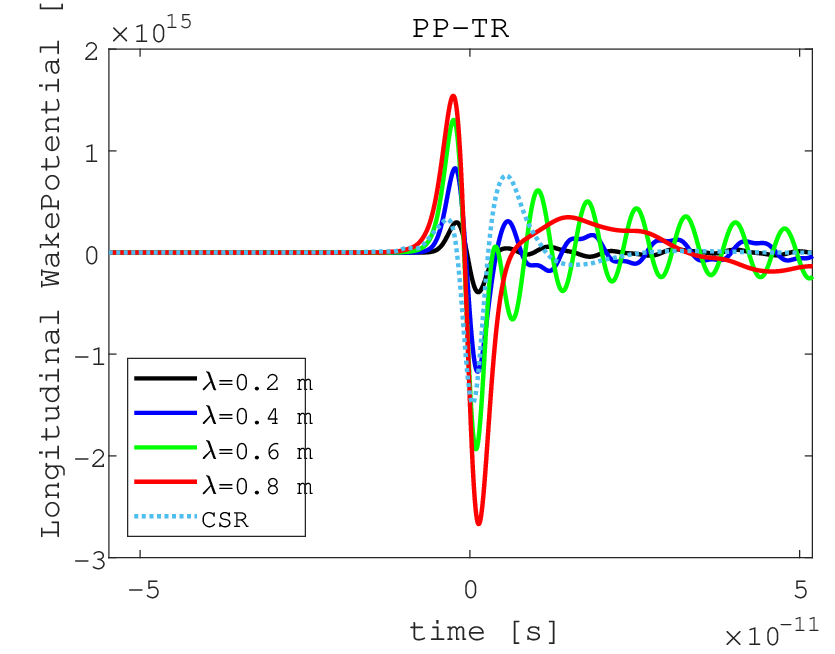}
  \caption{CWR wake potentials for a Gaussian bunch with $\sigma_z=0.5$ mm, obtained from the impedances in Fig.~\ref{fig:FS-PP-comparison}. Left: free space (FS-TR). Right: parallel-plates shielding (PP-TR). The legend indicates the wiggler period $\lambda_w$. In the right panel, the dashed cyan curve shows the CSR wake for the same bunch computed with the parallel-plates model.
  }
  \label{fig:WP-comparison}
\end{figure*}

STABLE has been successfully applied to study microwave instabilities induced by high-frequency resistive-wall impedance~\cite{WeiweiRW2024}. It is also well-suited for simulating instabilities driven by high-frequency CWR (or similar CSR) impedance. The program exclusively accepts short-bunch wake potential data as input, rather than direct impedance data. In this study, we employ a 0.5 mm rms Gaussian bunch length, which is less than one-tenth of the natural bunch length – sufficiently short to ensure accurate results for CWR instability analysis. Consequently, the CWR impedance data was converted into corresponding wake potentials. Figure~\ref{fig:FS-PP-comparison} presents the CWR impedance spectra, while Fig.~\ref{fig:WP-comparison} shows the corresponding wake potentials for various wiggler period lengths, based on the FS-TR and PP-TR models, respectively. For reference, we also include the CSR wake from the arc bending magnets, computed using the parallel-plates model~\cite{agoh2004calculation}. Here, the total wiggler length is fixed at $L_w = 76.8$ m and the peak field at $B_w = 1.6$ T, while the wiggler period length is varied. Consequently, the damping time remains constant, in accordance with the machine design requirement given by Eq.~\eqref{eq:DampingTime}. As the wiggler period length decreases (resulting in a decrease in $k_0$), the overall CWR impedance is clearly reduced in the regime where $k\ll k_0$. The computed wake potentials are then used as input for the STABLE tracking simulations. Each bunch consisted of 5 million macro-particles, with a fixed bin width of 0.1 ps. The single-bunch current was scanned from 0 to 3 mA. The beam was tracked for 50,000 turns, with statistics on bunch properties collected over the final 10,000 turns.
\begin{figure*}[htbp]
  \centering
    \includegraphics[width=\linewidth]{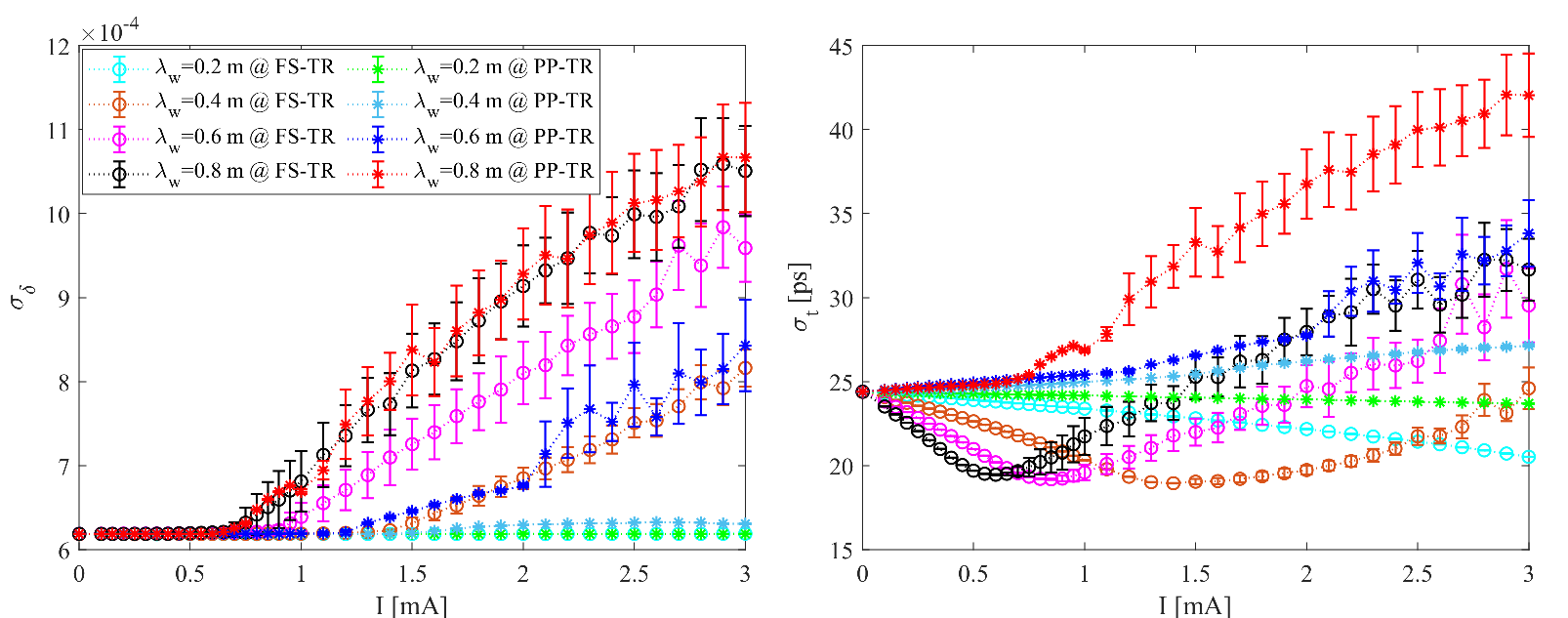}
  \caption{Comparison of tracking results between two models with different wiggler period lengths. The left plot shows energy spread as a function of bunch current, while the right plot displays rms bunch length. The legend indicates the period lengths and their corresponding calculation models.
  }
  \label{fig:trackingresults}
\end{figure*}

Figure~\ref{fig:trackingresults} plots the simulated STCF (1 GeV) energy spread and bunch length versus single-bunch current for CWR alone, comparing the FS-TR and PP-TR models. Threshold currents extracted from the tracking simulations are compared with analytical estimates in Table~\ref{table2}.
For a period length of 0.8 m in the current STCF design, both the analytical formula in Eq.~\eqref{eq:Ith_cwr_kc} and tracking simulations give a threshold of approximately 0.6 mA, which is significantly lower than the target current of 1.6 mA for the 1 GeV STCF. The energy spread results from both FS-TR and PP-TR modes shown in Fig.~\ref{fig:trackingresults} are relatively close. At 1.6 mA, the simulations predict a significant increase in energy spread. In the PP-TR model, the bunch length continuously increases with rising beam current. In contrast, the FS-TR model shows a non-monotonic behavior, with the bunch length initially decreasing and then increasing. This difference arises primarily because the FS models exhibit capacitive (deductive) impedance characteristics, whereas the PP models exhibit inductive characteristics, a distinction previously identified in~\cite{blednykh2023microwave} (see Fig.~17 therein). Considering that the actual situation is closer to the PP-TR model, the simulations predict that CWR will lead to significant increases in both energy spread and bunch length, which are expected to substantially impact the collision luminosity.
\begin{figure*}[htbp]
  \centering
    \includegraphics[width=\linewidth]{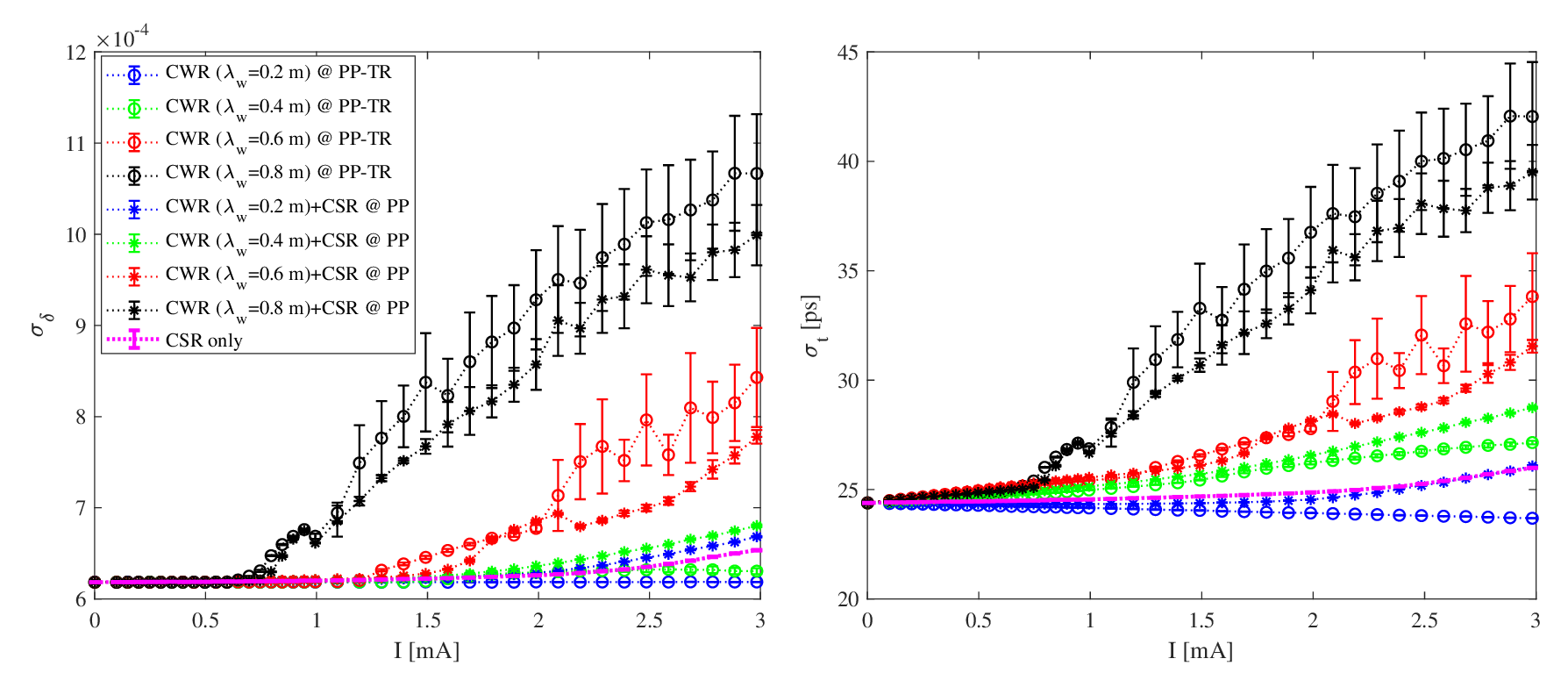}
  \caption{Comparison of tracking results for CWR (PP-TR) alone and for the combined CWR (PP-TR) + CSR (PP-SS). Left: energy spread versus single-bunch current. Right: rms bunch length versus single-bunch current.
  }
  \label{fig:trackingresults_CWRCSR}
\end{figure*}

\begin{table*}
\setlength{\tabcolsep}{5.0mm}
   \centering
   \caption{Comparison of CWR thresholds across computational models and methods for four wiggler period lengths.}
   \begin{tabular}{lcccc}
       \toprule
       \textbf{Period length} &{Particle tracking} &{Particle tracking} &{Particle tracking}  &{Using Eq.~\eqref{eq:Ith_cwr_kc}}\\ $\lambda_w$ [m]
         & @ PP-TR [mA] &  w/CSR @ PP [mA] & @ FS-TR [mA] & @ FS-SS [mA]\\
       \hline
           0.2 & 7.2 & 1.5 & 3.6  & 3.48\\
           0.4 & 1.4 & 1.3 & 1.4  & 1.26\\
           0.6 & 1.1 & 1.2 & 0.85 & 0.77\\
           0.8 & 0.6 & 0.7 & 0.6  & 0.59\\
       \hline
   \end{tabular}
   \label{table2}
\end{table*} 

The results in Table~\ref{table2} demonstrate that reducing the wiggler period length effectively increases the threshold current associated with CWR-induced instability. Moreover, the analytical predictions based on the FS-SS model Eq.~\eqref{eq:CWR_Impedance_FS-SS} closely match the numerical tracking results using the FS-TR model. This consistency confirms the validity of the analytical expression in Eq.~\eqref{eq:Ith_cwr_kc} reliably estimates the CWR threshold, supported by the fact that the condition $1<k_c\sigma_z\ll k_c$ is well satisfied in all cases. Furthermore, the PP-TR and FS-TR models give consistent threshold results for period lengths of 0.4 m and 0.8 m. However, for 0.6 m and 0.2 m, the PP-TR results are larger than the FS-TR results. Combining the tracking simulation results, as shown in Fig.~\ref{fig:trackingresults}, it can be concluded that by shortening the period length to 0.4 m, the threshold current reaches about 1.4 mA, slightly below the target current. In this case, the impact of CWR on luminosity would be limited and considered acceptable at 1 GeV for STCF.
\begin{figure}[htbp]
  \centering
    \includegraphics[width=\linewidth]{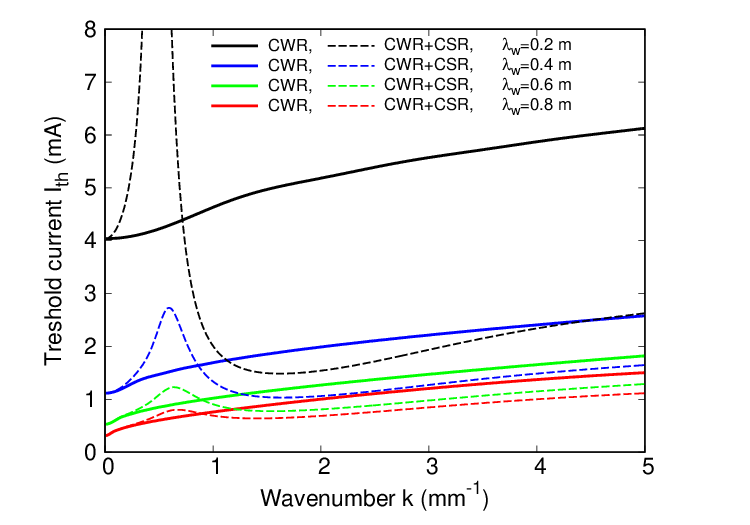}
  \caption{Threshold current as a function of wavenumber for STCF at 1 GeV. CWR and CSR impedances are modeled using the FS-TR and PP-SS formulations, respectively.
  }
  \label{fig:Ibth_analysis}
\end{figure}

We also examine the interplay between the CWR impedance and the CSR impedance from the arc bending magnets. Figure~\ref{fig:trackingresults_CWRCSR} contrasts simulations with PP-TR CWR alone against those including CSR, and the corresponding threshold currents are summarized in Table~\ref{table2}. With CSR included, the threshold changes little for $\lambda_w>0.2$ m but drops dramatically at $\lambda_w=0.2$ m. Thus, at $\lambda_w=0.2$ m the CSR impedance sets the instability threshold, whereas in the other cases the CWR impedance is the dominant limitation. To interpret these trends, we perform a dispersion-relation analysis based on Eq.~\eqref{eq:Ith-general}, using the computed FS-TR CWR impedance with and without the PP-SS CSR impedance (see Fig.~\ref{fig:Ibth_analysis}). We adopt the FS-TR model (rather than PP-TR) because its spectrum contains fewer narrow resonant spikes, which is preferable for dispersion-relation analyses of beam instability~\cite{stupakov2002beam}. As a practical procedure, we evaluate the threshold in the band $k\ge k_c(\text{CWR})$ and take the minimum in this range as the predicted current. With CWR alone, the analysis roughly reproduces the tracking thresholds obtained with the FS-TR model. With CWR+CSR, the CSR contribution grows as $\lambda_w$ decreases (while the CWR wake weakens as shown in Fig.~\ref{fig:WP-comparison}) and ultimately controls the threshold at $\lambda_w=0.2$ m. In summary, the simulation results in Figs.~\ref{fig:trackingresults} and \ref{fig:trackingresults_CWRCSR} agree with and are explained by the instability analysis.

\section{Discussion and Summary}\label{sec:summary}

The threshold current for longitudinal single-bunch instability in electron storage rings fundamentally scales with the parameter combination $\gamma\alpha_p\sigma_p^2\sigma_z$. In modern accelerator designs, particularly for light sources and circular colliders, this parameter tends to decrease significantly due to the push for ultra-low emittances to meet demanding experimental requirements. As a result, maintaining beam stability becomes increasingly challenging, and high-frequency impedance sources such as CSR and CWR emerge as critical considerations.

In this study, motivated by the design efforts for the STCF project, we investigated the theoretical aspects of the CSR and CWR impedances.
We showed that the CSR impedance can be compactly expressed as an equivalent effective impedance of Eq.~\eqref{eq:Zeff-CSR}, providing a convenient formulation for practical use.
For CWR, we developed several impedance models based on the analytical framework of~\cite{stupakov2016analytical}, with clearly defined applicability conditions to guide their usage.

We also revisited and extended the scaling laws for CWR-driven microwave instability, originally discussed in~\cite{wu2003impact}, to include the effects of damping wigglers and other lattice features specific to collider rings. The resulting expressions explicitly incorporate both wiggler and ring parameters, offering valuable insight for the optimization of storage ring designs.

As a case study, we applied the derived CWR impedance models to simulate the microwave instability for the STCF rings. The simulation results show good agreement with theoretical predictions, reinforcing the applicability and usefulness of the developed models in ring designs including wigglers. Moreover, the present framework is naturally connected to the beam measurements of the effects of the damping wigglers on microwave instability. For example, Ref.~\cite{brosi2018studies} reports that installing a single damping wiggler did not significantly change the instability threshold, whereas the associated reduction in damping time produced observable differences in the above-threshold radiation bursts. These observations provide an experimental context for the application and further validation of our models.

\section{Acknowledgments}
The author D. Zhou would like to thank G. Stupakov for valuable collaborations, from which the CWR impedance theories presented in this paper have greatly benefited. The author J.Y. Tang is supported by the National Natural Science Foundation of China (Project No. 12341501). The author T.L. He is supported by the National Natural Science Foundation of China (NSFC Grants No. 12375324 and No. 12105284) and the Fundamental Research Funds for the Central Universities (No. WK2310000127). The author L.H. Zhang is supported by the National Natural Science Foundation of China (Project No. 12405174). The authors also thank the Hefei Comprehensive National Science Center for the strong support on the STCF key technology research project.

\bibliographystyle{unsrt}
\bibliography{main.bib}
\end{document}